# Beyond Productivity Gaps: Temporal Patterns of Gender Differences in Scientific Knowledge Creation

Bili Zheng[1]; Chenyi Yang[1]; Jianhua Hou[1]

1 School of Information Management, Sun Yat-sen University, Guangzhou, 510006

**Abstract:** Gender inequality in scientific careers has been extensively documented through aggregate measures such as total publications and cumulative citations, yet the temporal dynamics underlying these disparities remain largely unexplored. Here we developed a multi-dimensional framework to examine gender differences in scientific knowledge creation through three complementary temporal dimensions: stability (consistency of performance over time), volatility (degree of year-to-year fluctuation), and persistence (ability to maintain high performance for extended periods). Using comprehensive bibliometric data from SciSciNet covering 62.5 million authors whose careers began between 1960-2010, we constructed knowledge creation capability measures that captured how scientists absorb knowledge from diverse sources and contribute to field advancement. We found that female scientists demonstrated significantly higher knowledge production stability (0.170 vs. 0.119 for males) while simultaneously exhibiting greater year-to-year volatility (6.606 vs. 6.228), revealing a striking paradox in career dynamics. Female scientists showed persistence advantages under moderate performance requirements but faced disadvantages under extreme criteria demanding sustained peak performance. However, these patterns varied substantially across disciplines, with female advantages strongest in humanities and social sciences while STEM fields show mixed results. Our temporal analysis revealed remarkable convergence over five decades, with stability gaps decreasing by 96% since the 1960s, though persistence differences remain stable. These findings challenged conventional bibliometric approaches by demonstrating that gender disparities operated primarily through complex temporal patterns rather than simple productivity gaps, with important implications for designing more equitable evaluation systems that recognized diverse pathways to scientific excellence.

**Keywords:** gender difference, gender inequality, knowledge creation, science of science

## 1. Introduction

Gender inequality in scientific careers has been extensively documented across multiple dimensions, from publication productivity and citation impact to collaboration patterns and career advancement (Huang et al., 2020; Modi et al., 2025). While considerable progress has been made in understanding the magnitude and scope of these disparities, most existing research focuses on aggregate measures of scientific output—total publications, cumulative citations, and career milestones—that capture the "what" but not the "how" of gender differences in scientific performance. A critical gap exists in understanding the temporal dynamics of scientific performance across gender groups. Scientific careers are inherently characterized by periods of varying intensity, breakthrough discoveries, and occasional setbacks (Way et al., 2017). How male and female scientists navigate these fluctuations, maintain consistency in their research contributions, and sustain high-level performance over extended periods remains largely unexplored. This temporal dimension is particularly important because current academic evaluation systems often emphasize sustained productivity and consistent impact, potentially disadvantaging groups that experience different career rhythm patterns.

Recent methodological advances in measuring knowledge production capabilities provide new opportunities to examine gender differences from a process-oriented perspective(Sebastian & Chen, 2021). Unlike traditional bibliometric indicators that count outputs (Bloom et al., 2020), knowledge creation measures capture how scientists create, integrate, and diffuse knowledge within the scientific ecosystem (Sebastian & Chen, 2021; Uzzi et al., 2013) . This shift from quantity-focused to process-focused analysis opens new avenues for understanding gender disparities in science. Building on these developments, this study introduces a novel analytical framework that examines gender differences through three complementary temporal dimensions: stability (consistency of knowledge production over time), volatility (degree of year-to-year fluctuation), and persistence (ability to maintain high performance for extended periods). By focusing on the dynamics rather than the levels of scientific performance, we aim to reveal patterns that may be obscured by traditional aggregate measures.

Our research addresses three fundamental questions: (1) Do male and female scientists exhibit different patterns of knowledge production stability and volatility across their careers? (2) How do gender differences in knowledge production persistence vary across disciplines? (3) What implications do these temporal patterns have for understanding and addressing gender inequality in science?

This investigation contributes to the literature in several important ways. First, we introduce novel metrics that capture temporal dynamics of scientific performance, moving beyond static comparisons to examine how gender influences the rhythm and sustainability of knowledge creation. Second, we provide comprehensive sensitivity analyses across multiple parameter specifications to ensure robust findings. Third, our results have practical implications for designing more equitable evaluation systems that account for different patterns of scientific contribution.

## 2. Literature review
### 2.1 Gender disparities in scientific performance: From aggregate measures to dynamic patterns

The study of gender differences in scientific careers has evolved through distinct phases, each characterized by different methodological approaches and theoretical perspectives. Early investigations primarily documented aggregate differences in productivity and impact, establishing what Cole and Zuckerman (1984) termed the "productivity puzzle"—the persistent finding that male scientists outperform female scientists on traditional bibliometric measures (Cole & Zuckerman, 1984). This foundational work, while crucial for documenting the scope of gender inequality, operated under an implicit assumption that scientific performance could be adequately captured through cumulative measures such as total publications and citation counts (Martin et al., 2025; Xie & Shauman, 1998).

However, this aggregate approach faces several limitations when applied to understanding gender differences. First, it treats scientific careers as static entities rather than dynamic processes unfolding over time (Way et al., 2017). Scientific productivity is inherently variable, with researchers experiencing periods of high output, breakthrough discoveries, and temporary setbacks (Sinatra et al., 2016). Second, aggregate measures may mask important differences in the temporal patterns through which male and female scientists achieve their career outcomes (Petersen et al., 2011). A scientist who maintains steady output over decades may achieve similar cumulative productivity to one who experiences periods of intense activity followed by relative quiet, yet these patterns have different implications for career evaluation and institutional support (Liu et al., 2021).

Contemporary research has begun to recognize these limitations, leading to calls for more sophisticated approaches to measuring scientific performance. Nielsen et al. (2017) demonstrated that gender diversity enhances research quality and innovation, suggesting that different approaches to knowledge creation may be equally valuable despite producing different aggregate outcomes. This perspective implies that understanding how gender influences the dynamics of knowledge production—rather than simply its cumulative results—is essential for developing more complete theories of scientific careers and more equitable evaluation systems (Nielsen et al., 2017).

**2.2 Temporal dynamics in scientific careers: The missing gender dimension**

Research on career dynamics has established that scientific productivity and impact follow complex temporal patterns that vary significantly across individuals, fields, and institutional contexts. The traditional model of scientific careers—characterized by early rapid growth followed by steady decline—has been challenged by empirical evidence showing much more diverse trajectory types (Sinatra et al., 2016; Way et al., 2017). These findings highlight the importance of understanding careers as dynamic processes rather than linear progressions, yet gender differences in these temporal patterns remain largely unexplored (Huang et al., 2020).

The few studies that have examined gender differences in career dynamics focus primarily on productivity measures and yield mixed results. Some research suggests that female scientists exhibit more consistent publication patterns over time, while male scientists show greater variability (Gao et al., 2021; Thelwall, 2020). However, these findings are based on simple productivity counts and do not consider the quality or impact dimensions of scientific output. Moreover, the mechanisms driving these differences remain unclear, as existing studies have not systematically examined how gender-specific career constraints and opportunities might shape temporal patterns of performance (Ceci & Williams, 2011).

This gap is particularly significant because temporal patterns may be more sensitive to gender-specific factors than aggregate measures. Life course events such as childbearing and childcare responsibilities, which disproportionately affect female scientists, may create specific patterns of career interruption and recovery (Antecol et al., 2018; Mason et al., 2013). Similarly, institutional factors such as access to resources, collaboration opportunities, and evaluation systems may differentially influence how male and female scientists structure their research activities over time (Ginther & Kahn, 2004). Understanding these temporal dimensions is crucial for identifying the mechanisms through which gender inequality emerges and persists in scientific careers.

**2.3 Conceptualizing performance dynamics: Stability, volatility, and persistence**

The concepts of stability, volatility, and persistence—well-established in economics and finance for analyzing dynamic systems—offer valuable frameworks for understanding temporal patterns in scientific careers (Mandelbrot & Hudson, 2007). Stability refers to the degree of consistency in performance over time, reflecting a scientist's ability to maintain steady output and impact (Petersen et al., 2012). Volatility captures the degree of fluctuation or unpredictability in performance, indicating how much a scientist's output varies from year to year (Radicchi et al., 2008). Persistence measures the ability to maintain high levels of performance for extended periods, reflecting sustained excellence rather than sporadic achievement (Ioannidis et al., 2014). These concepts are particularly relevant for understanding gender differences in scientific careers because they capture different aspects of how scientists navigate the inherent uncertainties and constraints of research environments. High stability might indicate effective resource management and strategic planning, while high volatility might reflect either adaptive responses to changing opportunities or the impact of external constraints (Azoulay et al., 2019). Similarly, persistence might indicate sustained access to resources and institutional support, while episodic high performance might reflect different career strategies or constraints (Liu et al., 2021).

From a gender perspective, several factors might influence these temporal patterns differently for male and female scientists. Research on work-life balance suggests that female scientists face greater challenges in maintaining consistent research focus due to family responsibilities (Antecol et al., 2018). Conversely, institutional research on collaboration patterns indicates that female scientists may develop more stable and long-term research relationships, potentially supporting more consistent performance over time (Araújo et al., 2017; van der Wal et al., 2021). However, systematic empirical investigation of how these factors translate into different patterns of stability, volatility, and persistence across gender groups is lacking (Abramo et al., 2021).

**2.4 Methodological innovations and knowledge creation measurement**

Recent advances in computational methods and large-scale data analysis have enabled more sophisticated approaches to measuring scientific performance that go beyond traditional bibliometric indicators (Lin et al., 2023). These innovations include network-based measures of collaboration and influence, semantic analysis of research content, and process-oriented measures that capture how scientists create and diffuse knowledge within their fields (Fortunato et al., 2018). Such approaches offer opportunities to examine the qualitative aspects of scientific contribution rather than focusing solely on quantitative outputs.

Knowledge creation capability measures, which assess how scientists absorb knowledge from diverse sources and contribute to the advancement of their fields, represent a particularly promising development for gender research (Holsapple & Joshi, 2002; Sebastian & Chen, 2021). Unlike traditional productivity measures that count outputs, these approaches capture the processes through which scientists create new knowledge and influence subsequent research (Grant, 1996). This process orientation is important because it may reveal gender differences that are obscured by aggregate measures—for instance, differences in research strategies, collaboration patterns, or approaches to knowledge integration (Kwiek & Roszka, 2022).

However, most applications of these advanced measurement approaches have not systematically examined gender differences, and those that have done so focus primarily on static comparisons rather than temporal dynamics (Holman et al., 2018). This represents a significant missed opportunity, as the combination of process-oriented measurement with temporal analysis offers the potential to reveal new insights into how gender influences scientific careers (Dion et al., 2018). Moreover, the temporal dimension is crucial for understanding the mechanisms through which gender differences emerge and evolve over time (Caplar et al., 2017).

**2.5 Research gaps and theoretical framework**

Our review reveals several critical gaps in the existing literature. First, while extensive research documents gender differences in aggregate scientific outcomes, little attention has been paid to temporal patterns of performance dynamics. Second, existing studies of career trajectories largely ignore gender differences or focus only on simple productivity measures. Third, the concepts of stability, volatility, and persistence—central to understanding dynamic systems—have not been systematically applied to examine gender differences in scientific careers. Fourth, recent advances in knowledge creation measurement have not been integrated with temporal analysis to examine gender differences in the dynamics of knowledge creation.

This study addresses these gaps by developing and applying novel measures of knowledge production dynamics to examine gender differences across multiple temporal dimensions. Our theoretical framework builds on three key insights from the literature: (1) scientific careers are inherently dynamic processes that cannot be adequately understood through aggregate measures alone, (2) gender differences may be more pronounced in temporal patterns than in cumulative outcomes, and (3) process-oriented measures of knowledge creation offer more nuanced insights into scientific contribution than traditional bibliometric indicators.

Based on this framework, we develop several hypotheses about gender differences in knowledge production patterns. We expect that female scientists will exhibit different patterns of stability, volatility, and persistence compared to male scientists, reflecting the different constraints and opportunities they face throughout their careers. We also anticipate that these patterns will vary across disciplines and career stages, as different institutional contexts and life course factors influence how gender shapes scientific careers. Finally, we expect that process-oriented measures of knowledge creation will reveal gender differences that are not apparent in traditional productivity measures, providing new insights into the mechanisms through which gender inequality operates in scientific careers.

## 3. Data and method
### 3.1 Data source

To investigate gender differences in the temporal dynamics of scientific careers, we utilize comprehensive bibliometric data that enables longitudinal analysis of individual research trajectories and knowledge production patterns. Our analysis is based on the SciSciNet database, which provides large-scale, longitudinal data on scientific publications and author careers (Lin et al., 2023). We collected data on individual scientists whose careers began between 1960 and 2010, restricting our analysis to authors with career spans of at least 10 years. For gender identification, we employed Genderize.io, a widely-used name-based gender inference service that has been validated in previous bibliometric studies (Santamaría & Mihaljević, 2018). To ensure accuracy, we applied a confidence threshold of 85% for gender assignments, excluding authors with lower confidence scores to minimize classification errors that could affect our gender-based comparisons.

Our final dataset comprises 62,522,361 authors, of whom 42,493,491 (68.0%) are identified as male and 20,028,870 (32.0%) are identified as female. This gender distribution is consistent with documented patterns of women's participation in science over the studied time period (Larivière et al., 2013). The substantial size of our dataset enabled robust statistical analysis of gender differences across different career stages, disciplines, and temporal patterns. For each author in our dataset, we collected comprehensive information including publication records, citation data, collaboration networks, and career-stage indicators.

### 3.2 Knowledge production capability measurement

Building on the recombination-based view of scientific knowledge creation, we develop a framework for measuring knowledge creation capability (KCC) that captures both the diversity of knowledge sources scientists draw upon and the breadth of their knowledge diffusion impact.

3.2.1 knowledge source diversity detection

We construct co-citation networks where two papers are linked if they appear together in the reference list of at least one target paper. We employ Informap algorithm (Yu et al., 2023) for community detection, which has been shown to provide superior performance identifying meaningful knowledge clusters by optimizing the map equation that describes the flow of information in networks. For each scientist's publication records in year $t$, we identify the knowledge clusters represented in their reference patterns and calculate knowledge source entropy:

$$E_{\text{source}}(s,t) = -\sum_{i=1}^{k} p(s,i,t) \cdot \log_2\bigl(p(s,i,t)\bigr) \qquad (1)$$

where $k$ is the number of distinct knowledge clusters identified by Informap in scientist $s$'s reference network at time $t$, and $p(s,i,t)$ represents the proportion of references belonging to cluster $i$.

3.2.2 Knowledge diffusion impact detection

Similarly, we construct forward citation networks linking papers that cite common sources, applying Infomap to identify knowledge domains influenced by a scientist's work. The knowledge diffusion entropy captures how broadly a scientist's contributions spread across different knowledge areas:

$$E_{\text{diffusion}}(s,T) = -\sum_{j=1}^{m} q(s,i,T) \cdot \log_2\bigl(q(s,i,T)\bigr) \qquad (2)$$

where $m$ represents the number of knowledge clusters in the forward citation network, $T$ represents the citing year, and $q(s,i,T)$ indicates the proportion of citing papers belonging to cluster $j$.

3.2.3 Knowledge creation capability matrix

Building on the matrix-based approach for measuring dynamic knowledge creation, we construct a temporal framework that captures both immediate knowledge absorption and long-term knowledge diffusion patterns. For each scientist, we create a knowledge production matrix spanning their career timeline.

Let $E_{\text{source}}(s,t)$ denote the knowledge source entropy of papers published by scientist $s$ in year $t$, and $E_{\text{diffusion}}(s,T)$ present the knowledge diffusion entropy of citations received in year $T$ by papers published in year $t$ ($T \geq t$). To account for varying citation dynamics and ensure temporal comparability, we focus on knowledge diffusion received within a 10-year window following publication. The knowledge creation capability for scientist $s$ in year $t$ is calculated as:

$$KCC_t = E_{\text{source}}(s,t) + \sum_{T=t}^{T+10} E_{\text{diffusion}}(s,T) \tag{3}$$

### 3.3 Temporal dynamic indicators

Building on the $KCC$ foundation, we develop three complementary indicators to capture different aspects of knowledge creation dynamics over scientific careers.

3.3.1 knowledge creation stability (KCS)

Stability measures the consistency of knowledge creation capability over a scientist's career, reflecting their ability to maintain steady research performance despite external variability. We define stability as the inverse of the coefficient of variation:

$$KCS_s = 1 - \frac{\sigma KCC_{(s)}}{\mu KCC_{(s)}} \tag{4}$$

Where $\sigma KCC_{(s)}$ and $\mu KCC_{(s)}$ represent the standard deviation and mean of scientist $s$'s $KCC$ values across their career. Higher $KCS$ indicates more consistent knowledge creation pattern, while lower one suggests greater variability.

3.3.2 Knowledge creation volatility (KCV)

Volatility captures the degree of year-to-year fluctuation in knowledge creation capability, measuring how unpredictably a scientist's performance varies over time. We calculate volatility as the root mean square of successive differences:

$$KCV(s) = \sqrt{\frac{1}{\mathbb{T}-1} \sum_{t=2}^{\mathbb{T}} [KCC(s,t) - KCC(s,\mathbb{T}-1)]^2} \tag{5}$$

Where $\mathbb{T}$ represents the career length of scientist $s$. This measure captures the magnitude of typical year-to-year changes in knowledge creation capability, wither higher values indicating more volatile patterns. KCV complements KCS by focusing specifically on temporal transitions rather than overall dispersion. A scientist might have low overall variability (high KCS) but experience occasional sharp transitions (moderate KCV), or conversely might have consistent gradual changes (low KCV) while showing high overall variability (low KCS).

3.3.3 Know creation persistence (KCP)

Persistence measures a scientist`s ability to maintain high levels of knowledge production capability for extended periods, capturing sustained excellence tranter than sporadic achievement. Our measurement approach employs individual-relative thresholds to ensure comparability across scientists with different baseline performance levels. For each scientist, we define high-performance years as those exceeding their personal $P_{th}$ percentile of KCC values:

$$Highligt\ year = \{t: KCC\ (s,t) > P_\tau[KCC\ (s)]\} \qquad (6)$$

Where $P_\tau$ represents the $\tau_{th}$ percentile of the scientist`s historical KCC distribution. We examine multiple percentile thresholds $\tau$ in $\{70, 75, 80\}$ to ensure robustness. We identify consecutive highlight periods with varying duration requirements and gap tolerances. The parameter combinations are:

2-year persistence: Minimum 2 consecutive years, gap to tolerance $g$ in $\{0,1\}$.

3-year persistence: Minimum 3 consecutive years, gap to tolerance $g$ in $\{0,1,2\}$.

4-year persistence: Minimum 4 consecutive years, gap to tolerance $g$ in $\{0,1,2\}$.

The persistence indicator is calculated as:

$$KCP_{(s,\tau,d,g)} = \frac{\sum_{i=1}^{N} length(period_i)}{\mathbb{T}_s} \qquad (7)$$

Where $N$ is the number of qualifying consecutive periods under parameters $(\tau, d, g)$, and $period_i$ denotes the duration of the $i_{th}$ qualifying period. This framework generates 18 persistence measures, enabling comprehensive sensitivity analysis across different definitions of sustained highlight.

4. Results

**4.1 Overall gender difference analysis**

Our analysis revealed distinct gender patterns in knowledge creation dynamics across three complementary dimensions. Female scientists demonstrated significantly higher stability in their knowledge creation capabilities (mean KCS: 0.170 vs. 0.119 for males, t-test $p < 0.001$), yet paradoxically exhibited greater year-to-year volatility (mean KCV: 6.606 vs. 6.228 for males, $t = -192.39$, $p < 0.001$). This apparent contradiction reflected fundamentally difference career dynamics: While female scientists maintained more consistent long-term performance levels, they experienced more pronounced fluctuations between individual years, potentially reflecting the influence of life course events and institutional constraints on their research trajectories.

The persistence dimensions revealed the most nuanced gender differences, with patterns that depended critically on the stringency of performance requirements. Under moderate persistence criteria – such as maintaining highlight for 2-3 consecutive years with allowances for brief interruptions – female scientist consistently outperformed their male counterparts (e.g., P70_C2_G1: female mean = 0.265, male mean = 0.261, $t = -16.65$, $p < 0.001$). However, this advantage reversed under the most demanding persistence requirements, where sustained excellence over 4+ years without interruption favored male scientists (e.g., P80_C4_G0: female mean = 0.003, male mean = 0.007, $t = 78.24$, $p < 0.001$) (Figure 1). This pattern suggests that while female scientists excelled at maintaining consistent intermediate-level performance, male scientists are more likely to achieve the extreme persistence required for sustained peak performance over extended periods.

The relationship between stability and volatility provided further insight into these gender-differentiated patterns. Despite their theoretical opposition, these measures showed only weak negative correlation ($r = -0.015$, $p<0.001$), confirming that they captured distinct aspects of knowledge production dynamics. This finding validated our multi-dimensional approach and suggested that high overall consistency (stability) and coexisted with significant year-to-year fluctuations (volatility). For female scientists, this combination may reflect adaptive strategies for managing competing demands throughout their careers, maintaining research productivity while navigating institutional barriers and life transitions that create temporary disruptions but do not derail long-term research trajectories.

These findings challenged conventional bibliometric approaches that focus on aggregate career outcomes. The temporal dynamics we observed suggested that gender differences in scientific careers were not simply matters of total productivity or impact, but reflected fundamentally different patterns of knowledge creation over time. Female scientists appeared to optimize for consistency and resilience, while male scientists more often achieved the extreme persistence patterns typically rewarded by academic evaluation systems.

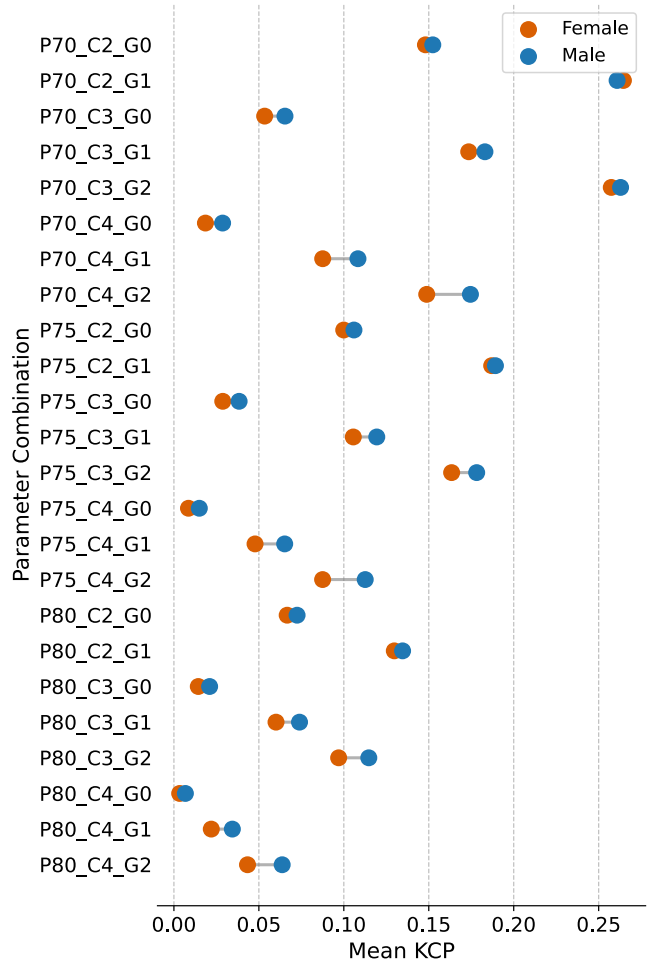

Figure 1. Gender difference in knowledge creation persistence by parameter.

### 4.2 Cross-disciplinary comparative analysis

The gender differences in knowledge creation dynamics exhibited variation across fields, revealing how institutional structures and research cultures shaped scientific careers differently for male and female scientists. Our analysis across 19 fields uncovered three distinct patterns that challenge assumptions about universal gender effects in science.

STEM fields demonstrated the most pronounced and consistent gender advantages for female scientists in stability, but reveaedl complex patterns in persistence. In disciplines such as Biology (female KCS: 0.279 vs. male: 0.219), Chemistry (0.254 vs. 0.189), Medicine (0.166 vs. 0.132), and Materials Science (0.224 vs. 0.154), female scientists consistently exhibited superior stability in knowledge production (Figure 2). However, the persistence revealed a more nuanced picture. While female scientists maintained advantages in fields like Physics (KCP: 0.178 vs. 0.173), Materials Science (0.167 vs. 0.166), and Environmental Science (0.164 vs. 0.168), they showed comparable or slightly lower persistence in traditional STEM fields like Biology (0.151 vs. 0.160), Chemistry (0.156 vs. 0.161), and Mathematics (0.159 vs. 0.162).

This pattern suggested that female scientists in STEM achieved superior long-term consistency through different mechanisms than sustained peak performance periods (Figure 3).

Humanities and social sciences revealed the most dramatic gender advantages across all three dimensions. In these fields, female scientists not only demonstrated superior stability—with both genders struggling but males facing particularly severe challenges in Art (female KCS: -0.818 vs. male: -0.949), History (-0.640 vs. -0.846), and Philosophy (-0.640 vs. -0.700)—but also showed substantially higher persistence. The persistence advantages were noticeable: Philosophy (female KCP: 0.087 vs. male: 0.074), History (0.080 vs. 0.066), Political Science (0.123 vs. 0.107), and Sociology (0.135 vs. 0.128).

The volatility dimension revealed an intriguing disciplinary paradox that inverts the stability findings. While female scientists demonstrated superior stability across most STEM fields, they simultaneously exhibited higher volatility in many of the same disciplines. Physics presented the most dramatic example (female KCV: 12.48 vs. male: 9.16), followed by Environmental Science (7.50 vs. 6.87) and Geography (7.75 vs. 7.18). This apparent contradiction—higher stability coexisting with higher volatility—suggested that female scientists in STEM achieved their superior long-term consistency through more pronounced year-to-year adjustments (Figure 4).

A clear disciplinary hierarchy emerged in persistence capabilities, with Physics and Materials Science at the top and Humanities at the bottom, but with consistent female advantages in humanities and mixed patterns in STEM. The persistence rankings revealed systematic differences across knowledge domains: Physics (0.178/0.173 for female/male), Materials Science (0.167/0.166), Environmental Science (0.164/0.168), and Computer Science (0.162/0.164) demonstrated the highest persistence levels, while humanities fields like Art (0.060/0.051), History (0.080/0.066), and Philosophy (0.087/0.074) showed much lower persistence. Notably, female scientists achieved their strongest relative advantages precisely in those fields with the lowest absolute persistence levels, suggesting that the mechanisms enabling sustained high performance may operate differently across fields.

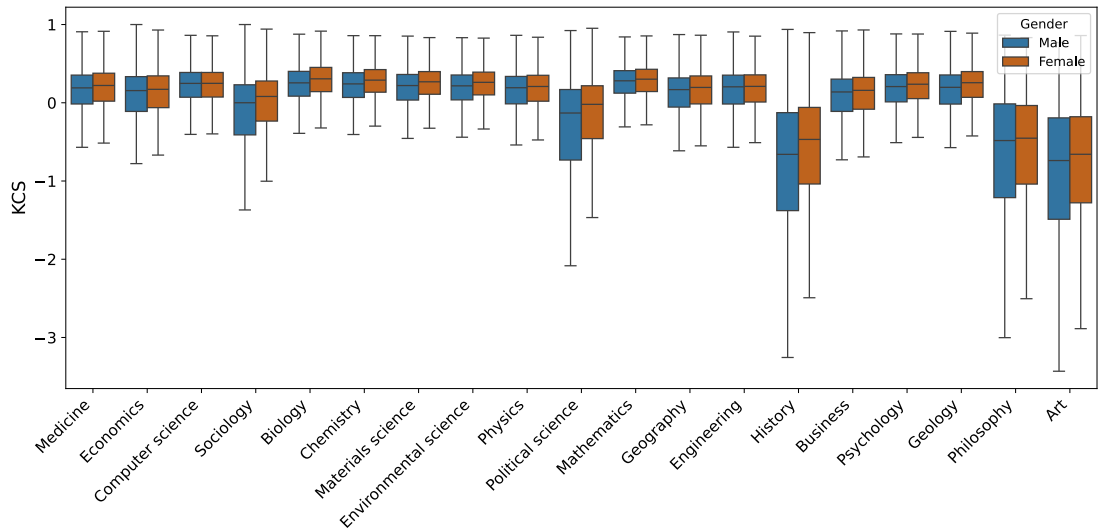

Figure 2. Gender differences in KCS across fields.

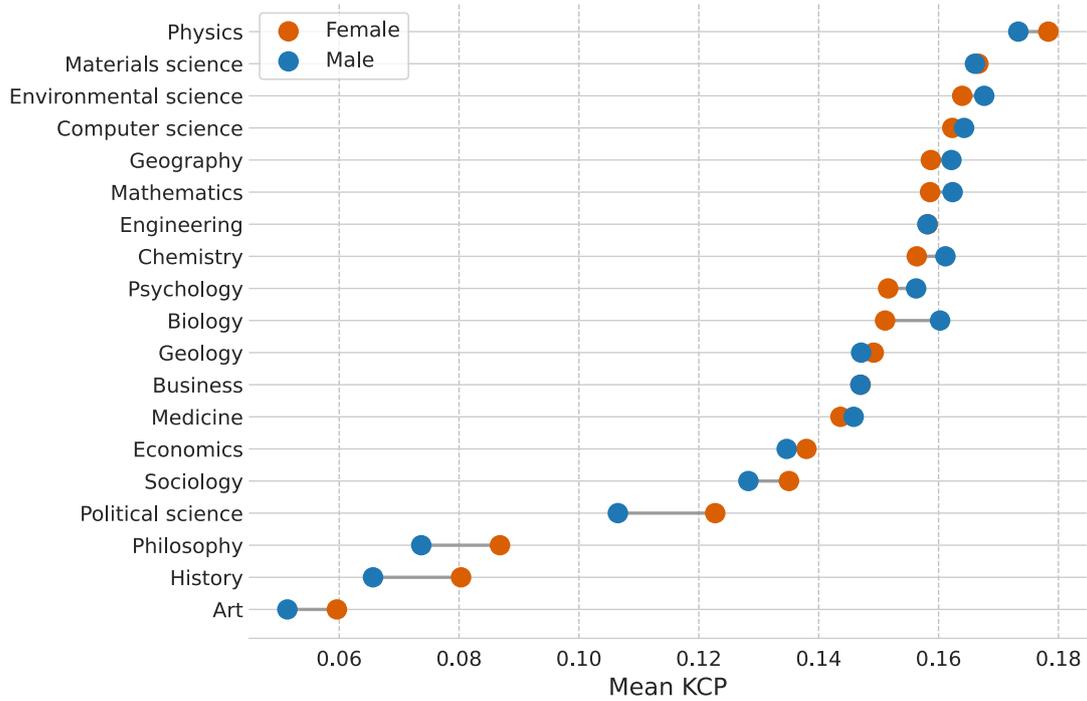

Figure 3. Gender differences in KCP for P70_C2_G0 across fields.

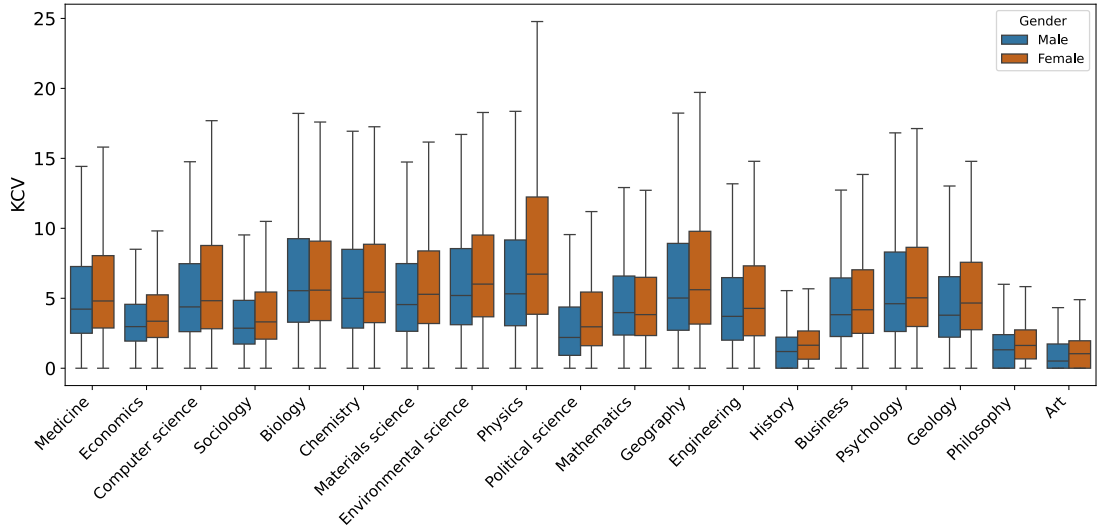

Figure 4. Gender differences in KCV across fields.

### 4.3 Temporal trends analysis

The temporal evolution of gender differences in knowledge creation dynamics revealed distinct patterns across five decades. Our cohort-based analysis, spanning from 1960 to 2010, demonstrated systematic changes in the magnitude and direction of gender disparities alongside overall improvements in scientific performance.

Knowledge creation stability showed a consistent convergence pattern across cohorts. The absolute gender gap in stability (male minus female) had decreased from -0.075 in the 1960-1964 cohort to -0.003 in the 2010+ cohort. The gender gap had similarly declined from 121% in the early 1960s to 1.3% in the most recent cohort. This convergence was particularly pronounced after the 1980-1984 cohort, when female scientists consistently outperformed male scientists in stability measures. Notably, both genders showed substantial improvements in absolute stability scored over time, with female scientists advancing from -0.062 to 0.227 and male scientists from -0.137 to 0.225 (Table 1).

Volatility patterns exhibited a more complex temporal trajectory with three distinct phases. Early cohorts (1960-1969) showed minimal gender differences in volatility, with gaps under 1.5%. The period from 1970-1999 demonstrated expanding gender differences, reaching a peak absolute gap of -0.275 (3.9% percentage gap) in the 1995-1999 cohort (Figure 5). The most recent cohorts showed a reversal of this pattern, with the 2010+ cohort exhibiting a positive gap of 0.073 (1.1%), indicating higher volatility among male scientists for the first time in our analysis period.

Persistence differences followed a distinct pattern compared to stability and volatility measures. The gender gap in persistence had remained relatively stable across cohorts, ranging from 2.5% to 5.4%, without the clear convergence pattern observed for stability. The 1960-1964 cohort showed a 5.4% gap, declining to 2.5% in the 1980-1984 cohort, then fluctuating between 2.8% and 4.3% across subsequent cohorts, ending at 5.2% for the 2010+ cohort. Despite this stable gender gap, both male and female scientists showed substantial improvements in absolute persistence levels, with female persistence increasing from 0.113 to 0.141 (25% improvement) and male persistence from 0.119 to 0.149 (25% improvement) (Figure 5).

Cross-cohort performance improvements were evident across all three dimensions for both genders. Female stability scores improved continuously from negative values (-0.062) in the 1960s to positive values (0.227) by 2010+. Male stability scores showed a similar trajectory from -0.137 to 0.225. Volatility patterns showed increases for both genders from the 1960s through the 1990s, followed by slight decreases in recent cohorts. Persistence measures demonstrated consistent improvements, with the most substantial gains occurring between the 1960s and 1980s cohorts (Table 1).

The temporal analysis revealed different convergence rates across the three knowledge production dimensions. While stability differences had decreased by approximately 96% over the study period, volatility differences peaked in the 1990s before beginning to reverse, and persistence differences had remained relatively constant in magnitude despite substantial absolute improvements for both genders. The 2010+ cohort represented a unique pattern where female scientists maintained advantages in stability while male scientists showed slightly higher volatility, yet the persistence gap remained similar to earlier decades.

Table 1. Gender gap in different cohorts.

| Cohort | KCS Gap (Male-Female) | KCS Gap (%) | KCV Gap (Male-Female) | KCV Gap (%) | KCP Gap (Male-Female) | KCP Gap (%) |
|---|---|---|---|---|---|---|
| 1960-1964 | -0.074962 | 121.296101 | -0.031248 | -0.875058 | 0.006047 | 5.369341 |
| 1965-1969 | -0.049565 | 138.152698 | -0.057151 | -1.335594 | 0.003959 | 3.098363 |
| 1970-1974 | -0.036596 | 674.561498 | -0.142781 | -2.872705 | 0.003882 | 2.809611 |
| 1975-1979 | -0.030773 | -97.226694 | -0.173455 | -3.109941 | 0.003696 | 2.562787 |
| 1980-1984 | -0.028743 | -41.970632 | -0.203623 | -3.287002 | 0.003762 | 2.513639 |
| 1985-1989 | -0.028468 | -27.011530 | -0.111519 | -1.69313 | 0.00437 | 2.841031 |
| 1990-1994 | -0.029684 | -21.582699 | -0.202941 | -2.945673 | 0.005056 | 3.268329 |
| 1995-1999 | -0.025658 | -15.198847 | -0.274819 | -3.929325 | 0.004558 | 2.988726 |
| 2000-2004 | -0.018800 | -9.738685 | -0.265286 | -3.883676 | 0.005394 | 3.609929 |
| 2005-2009 | -0.009488 | -4.412132 | -0.164866 | -2.519266 | 0.006177 | 4.280538 |
| 2010+ | -0.002961 | -1.301599 | 0.073306 | 1.135923 | 0.007309 | 5.173827 |

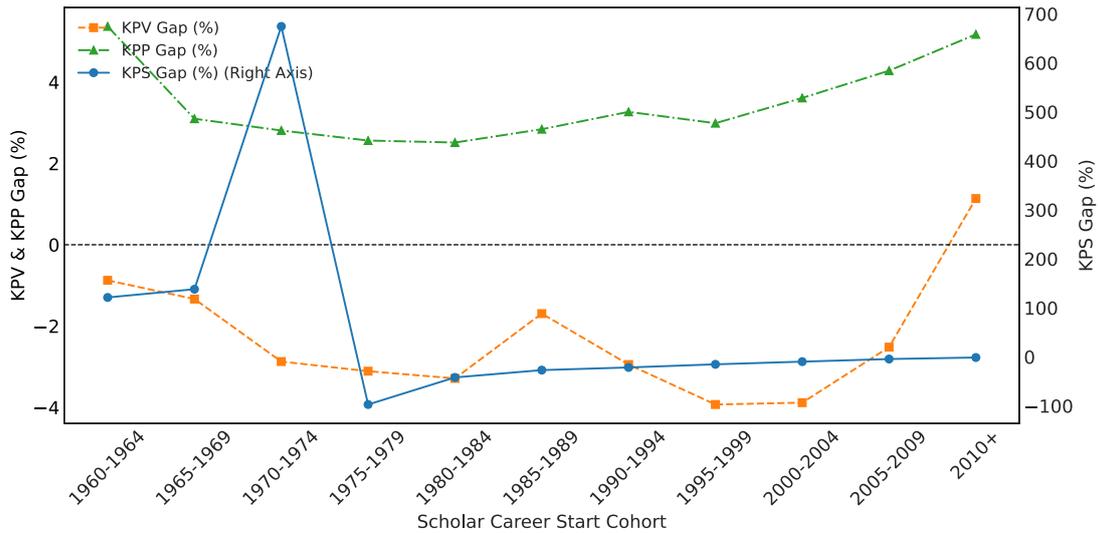

Figure 5. Trend of gender gap (%) by scholar cohort.

## 5. Discussion and conclusion

This study develops a multi-dimensional framework to examine gender differences in scientific knowledge creation through temporal dynamics rather than aggregate measures. We find that gender disparities in science operate primarily through complex patterns of stability, volatility, and persistence rather than simple productivity gaps. Female scientists demonstrate superior knowledge production stability (0.170 vs. 0.119) while simultaneously exhibiting higher volatility (6.606 vs. 6.228), with persistence advantages limited to moderate requirements but disadvantages under extreme criteria. These patterns vary substantially across disciplines and show remarkable convergence over five decades, with stability gaps decreasing by 96% since the 1960s while persistence differences remain stable. Our findings challenge conventional bibliometric approaches by revealing that understanding gender inequality in science requires examining how scientists structure their knowledge creation activities over time rather than focusing solely on cumulative outputs.

The most striking finding—that female scientists achieve higher stability while experiencing greater volatility—contradicts traditional assumptions about consistent performance patterns. This paradox extends beyond existing literature on gender differences in scientific careers, which has primarily focused on productivity gaps (Larivière et al., 2013) or citation disparities (Huang et al., 2020). Our results suggest that female scientists employ fundamentally different career optimization strategies, prioritizing long-term consistency over short-term smoothness in ways that previous aggregate measures could not detect. This pattern resonates with research on adaptive career management (Antecol et al., 2018; Mason et al., 2013), where temporary disruptions are managed through strategic portfolio adjustments that preserve overall trajectories. The weak correlation between stability and volatility (r = -0.015) confirms these capture distinct career dimensions, supporting theoretical models that emphasize multiple pathways to scientific success (Forthmann et al., 2025; Sinatra et al., 2016). Unlike studies that interpret volatility as career instability (Clauset et al.; Li et al., 2025), our findings suggest it may represent adaptive responses to institutional constraints and competing demands throughout academic careers.

The dramatic disciplinary variation in gender differences reveals how institutional structures and research cultures create different opportunity spaces for knowledge creation. Our finding that female advantages in STEM fields are strongest for stability but mixed for persistence contrasts with existing literature suggesting uniform disadvantages for women in these disciplines (Delgado & Murray, 2023; Sheltzer & Smith, 2014). In humanities and social sciences, female advantages across all dimensions (stability, volatility, and persistence) diverge from research emphasizing male dominance in academic hierarchies (Lerchenmueller et al., 2019). This disciplinary hierarchy—where female advantages are strongest in fields with lowest absolute persistence levels—suggests that evaluation systems and collaboration patterns operate differently across knowledge domains (Fortunato et al., 2018). The observation that Physics shows both high absolute persistence and the most dramatic volatility paradox (12.48 vs. 9.16) exemplifies how disciplinary context moderates gender differences in ways that previous field-specific studies have not captured (Hagiopol & Leru, 2024).

The temporal convergence patterns provide a new perspective on the evolution of gender equality in science. The 96% reduction in stability gaps since the 1960s represents more substantial progress than documented in studies focused on representation (Holman et al., 2018; Nielsen et al., 2017) or leadership positions (Funk & Owen-Smith, 2016; Wuchty et al., 2007). However, the persistence of gender differences in extreme performance requirements (2.5-5.4% across cohorts) suggests that fundamental mechanisms enabling sustained peak achievement continue to operate differently for male and female scientists. This finding complicates narratives of linear progress toward gender equality (Blau et al., 2010) by revealing that convergence rates vary across performance dimensions. The recent reversal in volatility patterns, with male scientists showing higher volatility for the first time in the 2010+ cohort, indicates that younger generations may be experiencing different career dynamics than previous research on generational differences has suggested (Kwiek & Roszka, 2022).

These findings have important implications for science policy and institutional practices. Current evaluation systems that prioritize sustained peak performance over consistent contributions may systematically disadvantage the career trajectories that female scientists more commonly follow, extending concerns about gendered evaluation criteria (Helmer et al., 2017; Wennerås & Wold, 1997). The observation that female scientists excel under moderate persistence requirements but show disadvantages under extreme criteria suggests that tenure timelines and interruption policies fundamentally shape equity outcomes (Lee et al., 2023; Lopes, 2024). Unlike research advocating for universal accommodations (Abir-Am & Outram, 1987; Moore et al., 2006; Valian, 1999), our results indicate that disciplinary variation in gender differences may require field-specific approaches to evaluation reform. The finding that high stability can coexist with high volatility challenges assumptions about "ideal" career trajectories and supports calls for recognizing multiple pathways to scientific excellence (Jones, 2010; Yang et al., 2025).

Several limitations constrain our interpretation of these findings. Gender identification through name-based algorithms may introduce systematic biases, particularly for international scholars (Santamaría & Mihaljević, 2018). Our observational design cannot establish causal relationships between institutional changes and observed convergence patterns. Additionally, we do not control for potentially important confounding factors such as institutional affiliations, funding access, or family status that might mediate observed gender differences (Kahn & MacGarvie, 2016). The knowledge production capability measure, while theoretically grounded, represents one approach to quantifying scientific contribution and may not capture all relevant aspects of knowledge creation (Uzzi et al., 2013; Wang et al., 2013).

Future research should investigate the specific mechanisms driving the stability-volatility paradox through longitudinal tracking of individual scientists across career transitions (Liu et al., 2023). Cross-cultural validation in different academic systems would test the generalizability of our findings beyond the publications captured in our dataset (Sugimoto et al., 2015). Investigating the relationship between these temporal patterns and career outcomes such as funding success or leadership positions would clarify the practical implications of different knowledge production strategies (Milojević et al., 2018; Stephan, 2012). As AI and other technologies reshape research practices (Kitchin, 2014; Tshitoyan et al., 2019), understanding how these changes affect gender differences in knowledge creation becomes increasingly important for building inclusive scientific institutions.

Overall, our analysis fundamentally reframes gender differences in scientific careers from whether such differences exist to how they manifest across multiple dimensions of knowledge production dynamics. By revealing that gender disparities operate primarily through temporal patterns rather than aggregate outcomes, we provide new insights for understanding persistent inequalities in scientific careers and designing more equitable evaluation systems. The multi-dimensional framework demonstrates that female and male scientists follow different but potentially equally valuable pathways to scientific contribution, with implications for how institutions can support diverse approaches to knowledge creation and sustained excellence.